# Superparamagnetic Transition and Local Disorder in $CuFe_2O_4$ Nanoparticles.

*G.F. Goya*[1] *and H.R. Rechenberg*
Instituto de Física, Universidade de São Paulo
C.P. 66318, 05315-970 / SP São Paulo - Brazil.

ABSTRACT

We present X-ray diffraction (XRD), Mössbauer spectroscopy (MS) and d.c. magnetization measurements performed on ball-milled $CuFe_2O_4$ samples. The average particle size <d> was found to decrease to the nanometer range after t=15 min of milling. Room temperature Mössbauer data showed that the fraction of particles above the blocking temperature $T_B$ increases with milling time, and almost complete superparamagnetic samples are obtained for <d> = 7(2) nm. Magnetization measurements below $T_B$ suggest spin canting in milled samples. The values of saturation moment $\mu_S$ reveal that site populations are slightly affected by milling. Mössbauer resonant intensities are accounted for on the basis of local disorder of $Fe^{3+}$ environments, and the development of sample inhomogeneities of $Cu_xFe_{3-x}O_4$ composition.

Short Title: Superparamagnetic $CuFe_2O_4$ Nanoparticles

[1] Corresponding author. E-mail: goya@unizar.es

## INTRODUCTION

Macroscopic magnetic tunneling, magnetic relaxation, and spin canting are some of the phenomena recently found in nanosized materials, which have become an intense research field(1-4). An increasing fraction of these nanosized materials are being obtained by mechanical grinding, and the resulting phases have shown rather peculiar structural and magnetic properties(5-8). Relaxation effects in mixed ferrites have been studied(9-11) and recently, spin canting in maghemite nanoparticles induced by structural disorder have been reported(12).

The copper spinel structure has a cubic close-packed arrangement of the oxygen ions, with the $Cu^{2+}$ and $Fe^{3+}$ ions in two different kind of sites(13). These sites have tetrahedral and octahedral oxygen coordination (A- and B-sites respectively), so the resulting local symmetries of both sites are different. Additionally, there is an off-center displacement of the metallic ions, usually described in terms of a parameter *u*, which depends on the ionic radii of the transition metals involved. The local symmetry of A-site does not depend on the value of *u*, but B-sites are distorted by deviations of the ideal value *u*=3/8. The resulting ionic distribution in this kind of structure may be represented by

$$[Cu_{\delta}Fe_{1-\delta}]^{A}[Cu_{1-\delta}Fe_{1+\delta}]^{B}O_4 \qquad \text{( I )}$$

where $\delta$=0 and $\delta$ =1 stand for the inverse and normal cases, respectively. Although most spinel ferrites are cubic, $CuFe_2O_4$ can have tetragonal unit cell symmetry if cooled slowly from high temperatures, with values of $\delta$ between 0.1-0.2 (14-16). Attempts to explain the resulting ionic distribution from electrostatic (Madelung) energy calculations have led only to qualitative results(13). In $CuFe_2O_4$ spinel, the presence of $Cu^{2+}$ with orbitally degenerate ground state lead to expect Jahn-Teller interaction to be operative, modifying the local symmetry of the sites.

We present in this work Mössbauer spectroscopy, d.c. magnetization and x-ray diffraction (XRD) data on tetragonal $CuFe_2O_4$ spinel ferrite, ball milled using “low” energy (low rotational velocity) and short times (several minutes), to investigate the superparamagnetic (SPM) transition and the resulting ionic distribution after the grinding process.

## EXPERIMENTAL

Samples of $CuFe_2O_4$ spinel were prepared by standard solid state reaction, mixing stoichiometric amounts of $\alpha$-$Fe_2O_3$ and CuO (99.999%). The mixture was fired in air at 900 C for 72h, with intermediate grinding in agate mortar. Samples were cooled slowly (2 K/min) to avoid quenching of the high-temperature cubic spinel phase. Ball milling of the resulting $CuFe_2O_4$ was carried out in a planetary ball mill (Fristch Pulverisette 7) with hardened steel vials and balls. The ratio of masses was chosen as 10:1, and the milling intensity set as 500rpm. The process was interrupted at several intermediate times to take out small parts of the powder. Samples were labeled S0; S15; S45; S105; S200 and S300 accordingly to the total minutes of grinding time. XRD characterization was performed in a Philips PW-1140 diffractometer using Cu-K$\alpha$ radiation. Magnetization measurements were performed in a vibrating sample magnetometer at 300K and 4.2K, in fields up to 90 kOe. Mössbauer data was taken at 4.2 , 78 and 296 K in transmission geometry, using a $^{57}$Co source of about 50 mCi in constant acceleration mode. A non-linear least-squares program was used to fit the spectra to Lorentzian line shapes. Isomer shifts are referred to $\alpha$-Fe at 300K.

## RESULTS AND DISCUSSION

Sample S0, as prepared for the milling experiment, was found to be single phase $CuFe_2O_4$ spinel (tetragonal unit cell, space group $D_{4h}$ - I 41/amd), with lattice parameters $a_0 =$ 8.222(5) Å and $c_0 =$ 8.690(5) Å. The XRD patterns of pristine and milled $CuFe_2O_4$ (figure 1) show the broadening of the peaks with milling, although their positions do not change along the process. Average grain sizes <d> were estimated from Scherrer's equation, subtracting instrumental broadening from the experimental linewidth. The resulting values of <d> are shown in the inset of figure 1. It is observed that particle size drops abruptly to <d>=20(2) nm after the first 15 minutes of milling. Afterwards, <d> decreases monotonically to a minimum value <d>=7(2) nm.

Mössbauer spectrum of pristine $CuFe_2O_4$ (sample S0), shown in figure 2, is composed of two partially resolved magnetic sextets arising from $Fe^{3+}$ in tetrahedral (A) and octahedral (B) sites, as previously reported(17,18). The spectra of milled samples S15 to S105 could be fitted with two magnetic plus one quadrupolar signals (Table I and Figure 2). The hyperfine parameters of the two sextets vary slightly along the series, implying that both A and B local environments remain essentially unchanged. The increasing central doublet in figure 2 corresponds to the SPM transition of the single-domain particles having "effective" volumes with blocking temperatures $T_B$ < 296K, as measured within the time scale of Mössbauer spectroscopy, $\tau \sim 10^{-9}$ s. Figure 3 shows the evolution of blocked and SPM fractions versus milling time. The superparamagnetic $Fe^{3+}$ fraction amounts to less than 15% of the total resonant signal for S15 to S105 samples, all with average particle diameter <d> > 10 nm. After the next t=100 min of milling (i.e., sample S200) the SPM phase increases to 40%, and for S300 the fraction of SPM particles is 91%.

TABLE I:

Mössbauer Hyperfine Parameters taken at T = 296K.

H = hyperfine field, δ = isomer shift, Δ = quadrupole splitting, Γ = linewidth and I = spectral area of each Fe site. Errors are quoted between parenthesis.

| Sample | Parameters | H-1 | H-2 | P-1 | P-2 |
|---|---|---|---|---|---|
| S0 | H (T) | 51.0(1) | 48.3(1) | -- | -- |
| | δ (mm/s) | 0.36(1) | 0.26(1) | | |
| | Δ (mm/s) | -0.14(1) | -0.02(1) | | |
| | Γ (mm/s) | 0.45(1) | 0.46(1) | | |
| | I (%) | 43(2) | 57(2) | | |
| S15 | H (T) | 50.8(1) | 48.0(1) | | -- |
| | δ (mm/s) | 0.38(1) | 0.27(1) | 0.39(4) | |
| | Δ (mm/s) | -0.14(1) | -0.02(1) | 1.73(2) | |
| | Γ (mm/s) | 0.40(2) | 0.62(3) | 0.99(5) | |
| | I (%) | 24(4) | 68(4) | 8(1) | |
| S45 | H (T) | 50.8(1) | 48.0(1) | | -- |
| | δ (mm/s) | 0.37(1) | 0.28(1) | 0.41(4) | |
| | Δ (mm/s) | -0.14(2) | -0.02(4) | 1.59(7) | |
| | Γ (mm/s) | 0.39(3) | 0.62(2) | 1.02(8) | |
| | I (%) | 22(3) | 68(4) | 11(1) | |
| S105 | H (T) | 50.7(1) | 48.0(1) | | -- |
| | δ (mm/s) | 0.38(1) | 0.28(1) | 0.38(4) | |
| | Δ (mm/s) | -0.13(1) | -0.02(1) | 1.37(7) | |
| | Γ (mm/s) | 0.40(5) | 0.66(3) | 1.1(1) | |
| | I (%) | 18(3) | 67(4) | 14(2) | |
| S200 | H (T) | 47.8(1) | 44.3(5) | | |
| | δ (mm/s) | 0.31(2) | 0.28(1) | 0.31(1) | 0.29(3) |
| | Δ (mm/s) | -0.02(1) | -0.04(2) | 1.00(5) | 1.99(2) |
| | Γ (mm/s) | 0.57(7) | 0.98(8) | 0.69(5) | 0.75(5) |
| | I (%) | 18(4) | 40(6) | 27(5) | 12(5) |
| S300 | H (T) | -- | 43.4(1) | | |
| | δ (mm/s) | | 0.35(1) | 0.34(1) | 0.33(1) |
| | Δ (mm/s) | | -0.12(2) | 0.72(2) | 1.19(4) |
| | Γ (mm/s) | | 0.99(5) | 0.44(3) | 0.75(2) |
| | I (%) | | 8(2) | 25(8) | 65(8) |

This sudden increment in the SPM phase occurs concurrently with the "average" crossing of the <d> $\approx$ 10 nm particle size. This may be explained assuming that $d_C$=10 nm is approximately the SPM critical size of this phase, which is comparable to previous estimations in similar systems(7,19,20).

For S200 and S300 samples, the SPM doublet is the major signal. The fit of this central doublet needed two quadrupolar components as shown in Table I. This agrees with previous Mössbauer data on $^{57}$Fe-doped tenorite ($Cu_{0.99}Fe_{0.01}$)O (14), where isolated SPM particles of $CuFe_2O_4$ in a CuO matrix displayed two-component spectra. These two components can be related to a) different SPM signals from A- and B-sites of spinel, or b) surface and "bulk" contributions from the SPM particles. In the first case, the intensity ratio $R_P$=I(P2)/I(P1) is expected to be related to the relative A- and B-site populations (inversion degree), irrespective of particle size. On the other hand, a higher surface/volume ratio for smaller particle sizes will lead to an increment of $R_P$. The increase of the intensities ratio $R_P$ from S200 to S300 supports the latter assignment. Attempts to fitting the spectra with two quadrupolar signals did not improve the $\chi^2$ value significantly in samples with less than 15% of SPM particles, since this small amount prevents clear separation between the doublets and the central part of the magnetic sextets. The differences observed in IS and $\Delta$Q values between reference (14) and the present data are probably originated in the different particle interactions of the CuO and $CuFe_2O_4$ matrix, respectively. Unfortunately there is not, to our knowledge, any reliable model capable of relating experimental relaxation spectra with the physical parameters of a volume distribution of SPM particles, so further analysis was not performed. Defining the blocking temperature $T_B$ as the temperature where both SPM and ordered particles have equal resonant intensities, we see in figure 3 that $T_B \approx$ 296K for S200 sample.

TABLE II:

Mössbauer Hyperfine Parameters for sample S300 taken at T = 4.2 K. H=hyperfine field, Δ = quadrupole splitting, δ = isomer shift, Γ=linewidth and I=spectral area. Errors are quoted between parenthesis.

| Sample | Parameters | H-1 | H-2 | H-3 |
|---|---|---|---|---|
| S300 | H (T) | 52.8(5) | 50.2(4) | 46.2(6) |
| | δ (mm/s) | 0.49(2) | 0.43(2) | 0.40(3) |
| | Δ (mm/s) | -0.07(2) | -0.02(2) | -0.09(2) |
| | Γ (mm/s) | 0.45(2) | 0.59(3) | 0.74(6) |
| | I (%) | 25(4) | 45(4) | 31(4) |

In order to check the SPM origin of the central doublet, we performed Mössbauer measurements at 78 K on S200 and S300 samples (Figures 4a. and b.). At this temperature only magnetically split components are observed, as expected for particles below $T_B$. Some residual relaxation effects are, however, still present as evidenced by the background curvature of both spectra. This curvature is more pronounced for S300 than for S200, as expected from the smaller particle size. In figure 4c, the Mössbauer spectrum of S300 at T = 4.2 is shown, where relaxation effects are absent. The hyperfine parameters of H1 and H2 signals (Table II) correspond to A and B sites of the spinel at this temperature, slightly reduced by the small particle size(21). A third sextet is observed, centered at a lower field of H3=46.2 T and asymmetrically broadened, indicating a distribution of hyperfine fields. This may be due to the formation of the solid solution $Cu_xFe_{3-x}O_4$ with an extended range of x, where the different $Fe^{3+}$ environments result in the observed hyperfine field distribution(22). It has been reported that ball milling in a closed container transforms $\alpha$-$Fe_2O_3$ into $Fe_3O_4$(23). The formation of sample areas with $Cu_xFe_{3-x}O_4$ stoichiometries, after 300 minutes of milling, seems to be related to the beginning of this Fe reduction. The broadening of the XRD peaks in milled samples prevented the confirmation of $Cu_xFe_{3-x}O_4$ phase, due to the overlap with the corresponding lines of $CuFe_2O_4$.

Figure 5 shows the normalized magnetization of sample S300, plotted against reduced field H/T, at 300 and 4.2 K. The curves clearly indicates that sample S300 is below the SPM transition at T=4.2 K, for the measuring times $\tau_m \approx 100$ s involved in magnetic measurements. Table III shows the values of saturation magnetization $M_S$, coercive force $H_C$, reduced remanence ($M_R/M_S$) and saturation moment $\mu_S$ obtained in an external field H= 90 kOe, at 4.2 and 300 K. It is observed that $M_S$ decrease from S0 to S300, both in the SPM and blocked states. This may be due to small contamination with $\alpha$-$Fe_2O_3$ , and/or a canted spin structure resulting from the milling process. The absence of saturation for sample S300 (see figure 5), even at fields H=90 kOe, supports the latter assumption.

TABLE III.

Magnetic Data for Samples S0 and S300, at 4.2 and 300 K.
Saturation Magnetization ($M_S$), Reduced Remanence ($M_R/M_S$),
Coercivity ($H_C$) and Saturation Moment ($\mu_S$)

| SAMPLE | T (K) | $M_S$ (emu/g) | $\mu_S$ ($\mu_B$/f.u.) | $M_R/M_S$ | $H_C$ (kOe) |
|---|---|---|---|---|---|
| S0 | 300 | 32.8(3) | 1.38(3) | 0.51(5) | 0.75(2) |
| | 4.2 | 33.4(3) | 1.41(3) | 0.52(4) | 0.83(2) |
| S300 | 300 | 26.3(3) | 1.11(3) | 0.12(5) | 0.15(2) |
| | 4.2 | 29.5(3) | 1.25(3) | 0.50(4) | 1.11(2) |

The resulting particle size distribution after milling leads to non-zero coercive force at 300 K for sample S300 (Table III), due to a fraction of particles in the blocked state. At T=4.2 K the value of $H_C$ increases about 7 times, as seen in the inset of figure 5, since most particles are below $T_B$. The remanence values found for bulk $CuFe_2O_4$ are similar to previous data from polycrystalline samples(24). These values are somewhat lower than the calculated values 0.8-0.9 for ferrites with cubic crystal structures and different preferred directions of the magnetization(25). The similar values of $M_R/M_S$ measured below $T_B$ for S0 and S300 samples indicate that the milling process does not introduce substantial changes in crystal and/or shape anisotropies. It can be further noticed that, at 4.2K, the $M_R/M_S$ ratio is very close to the ideal 0.5 value characteristic of single-domain noninteracting particles. At T=300 K, the large reduction of this ratio from S0 to S300 samples adds evidence to the significant SPM behavior. The non-zero remanence value for S300 at room temperature is associated with the fraction of particle sizes in the blocked state.

In $CuFe_2O_4$ , the total saturation moment per formula unit, $\mu_S$, is related with the inversion degree $\delta$ by $\mu_S = \mu_{Cu} + 2\delta\ (\mu_{Fe} - \mu_{Cu})$, where $\mu_{Fe}$ and $\mu_{Cu}$ are the $Fe^{3+}$ and $Cu^{2+}$ moments, respectively. Taking $\mu_{Fe} = 5\ \mu_B$ and $\mu_{Cu} = 1\ \mu_B$, we have $\mu_S = (8\ \delta + 1)\ \mu_B$, and thus small changes in $\delta$ leads to large increments in $\mu_S$ (i.e., for the exchange of one B-site $Cu^{2+}$ by one A-site $Fe^{3+}$ , and viceversa, we have $\delta$=0.125, and the saturation moment rises from 1 to 2 $\mu_B$ ). Magnetic data from Table III show that $\mu_S$ values decrease about 10% after 300 minutes of ball milling. As discussed above, spin canting and/or $Cu_xFe_{3-x}O_4$ formation can account for this reduction, meaning that site populations in $CuFe_2O_4$ are not appreciably affected by milling.

From the above discussion we infer that the change in the relative intensities of Mössbauer lines, along the milling series, is not associated with changes in site populations. As discussed in reference(17) the use of Lorentzian lineshape to fit the magnetic signals leads to erroneous estimations of A and B site occupancies, attributed to anomalous lineshape of B-site pattern and the poor resolution between A and B signals. For our milled samples, it is observed that longer grinding times increase the single intensity I(H2) from 57 % to 79 % of the total magnetic signal. From figure 2 and Table I it is clear that the intensity increment of H2 can hardly be attributed to resolution effects. The ratio of A to B Mössbauer intensities, $R_H$ = I(H1)/I(H2), is related to the inversion degree by $R_H = (1+\delta)/(1-\delta)$. The fit to the spectra using Lorentzian lineshapes gives values of $R_H$ ranging from 0.75 to 0.27, for S0 and S105 samples respectively. These values are clearly incorrect, since a $R_H < 1$ value has no physical meaning.

It is known that $CuFe_2O_4$ spinel shows a marked Jahn-Teller effect, arising from the doubly degenerate $E_g$- type ground state of the $Cu^{2+}$ ion. These local distortions give rise to fluctuations of the superexchange interactions that broadens the resonant line. From Table I it can be seen that $\Gamma_1 < \Gamma_2$ in all milled samples. Additionally, the increase of $\Gamma_2$ with t indicates that A site is more affected by ionic disorder than B site. This can be understood assuming that, as a result of the milling process, a fraction of B sites are distorted to a cubic-like symmetry, thus contributing to the resonant line of the A site. These changes in local symmetry induced by structural disorder might be driven by similar mechanisms to those determining the cubic $CuFe_2O_4$ phase when quenching from T>900 °C, by freezing the high-temperature disordered phase. However, we cannot discard that an additional source of fluctuations in the hyperfine field might be an incipient amount of inhomogeneities of composition $Cu_xFe_{3-x}O_4$ discussed above, which becomes detectable in S300 below $T_B$.

In summary, we have studied the structural and magnetic properties of nanosized particles of $CuFe_2O_4$ spinel above and below the superparamagnetic transition. We found that several minutes of mechanical grinding are enough to reduce the average grain size to the nanometer range. After 300 minutes of mechanical grinding, the fraction of SPM spinel is almost 100 per cent. The magnetic hyperfine fields, coercive force and reduced remanence measured below $T_B$ showed that, at this temperature, samples are in the blocked state. No evidence of new phases was found up to 200 minutes of milling, but for 300 minutes a new $Fe^{3+}$ site is observed, assigned to $Fe^{3+}$ in $Cu_xFe_{3-x}O_4$. Magnetic data below the blocking temperature indicated that milling does not modify the inversion degree substantially. The observed decrease in the magnetic moment with milling, and the absence of saturation at 90 kOe, suggest that spin canting might occur. The decrease ratio $R_H$ of Mössbauer intensities with milling is attributed to a) distortions of the local environments of Fe sites by milling-induced disorder, and/or b) the formation of sample inhomogeneities with composition $Cu_xFe_{3-x}O_4$.

ACKNOWLEDGMENTS

One of us (G.F.G.) thanks financial support from FAPESP through a post-doctoral fellowship.

Figure Captions

Figure 1. X-ray diffraction data for $CuFe_2O_4$ samples a) *as prepared* (S0) and b) mechanically ground for 300 min (S300). The inset shows the average grain size <d> *vs.* milling time, as estimated from Scherrer's equation. For S0 sample, <d> ~ 200 nm lays out of the scale.

Figure 2. Room-temperature Mössbauer spectra of samples milled for different times. Solid line is the best fit to experimental data (solid circles) whereas dotted lines are the magnetic and quadrupolar subespectra.

Figure 3 Evolution of the total magnetic and quadrupolar Mössbauer resonant intensities with milling time. Solid circles correspond to paramagnetic signal, whereas solid up triangles to magnetic signal. Lines are a guide to the eye.

Figure 4. Mössbauer spectra of a) sample S200 and b) sample S300 at T=78K. c) Sample S300 measured at 4.2 K. Dotted lines: magnetic subespectra. Solid line: best fit to experimental data.

Figure 5. Magnetization curves *vs.* reduced field (H/T), taken at 300 and 4.2 K, of sample milled 300 minutes. The maximum field is H=90 kOe. The inset shows the low-field region.

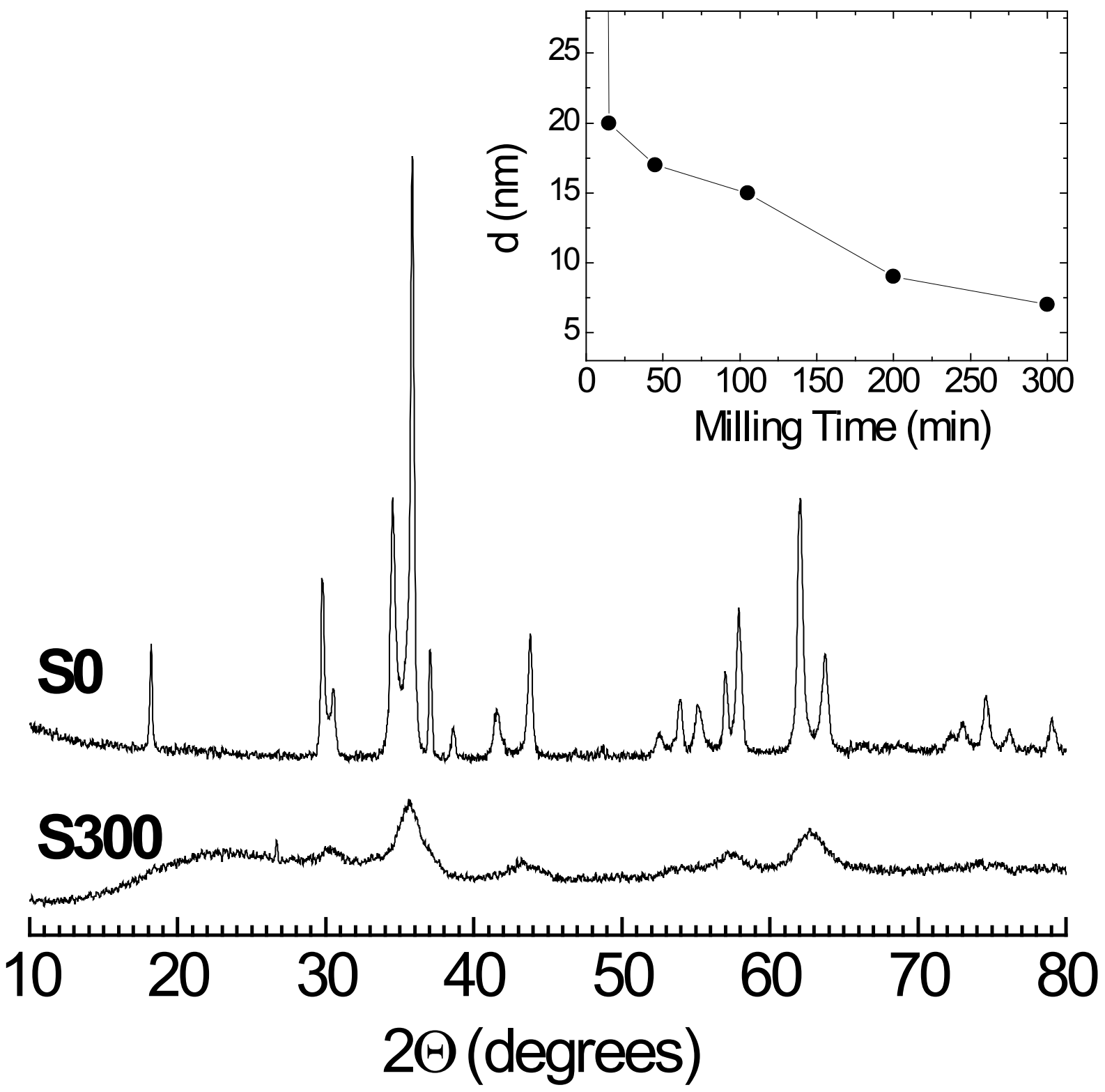
S0
S300
10 20 30 40 50 60 70 80
2Θ (degrees)
d (nm)
5 10 15 20 25
0 50 100 150 200 250 300
Milling Time (min)

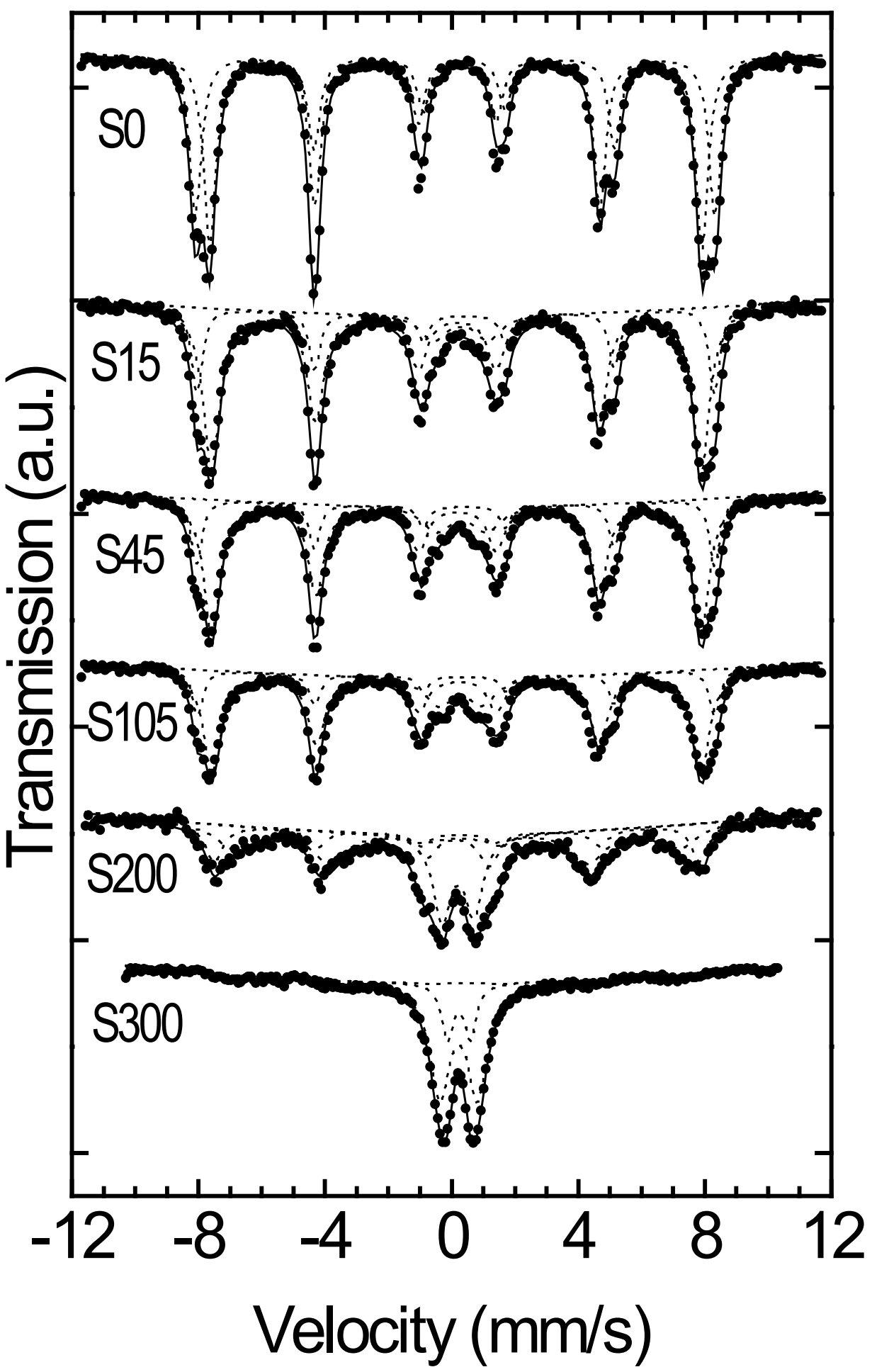

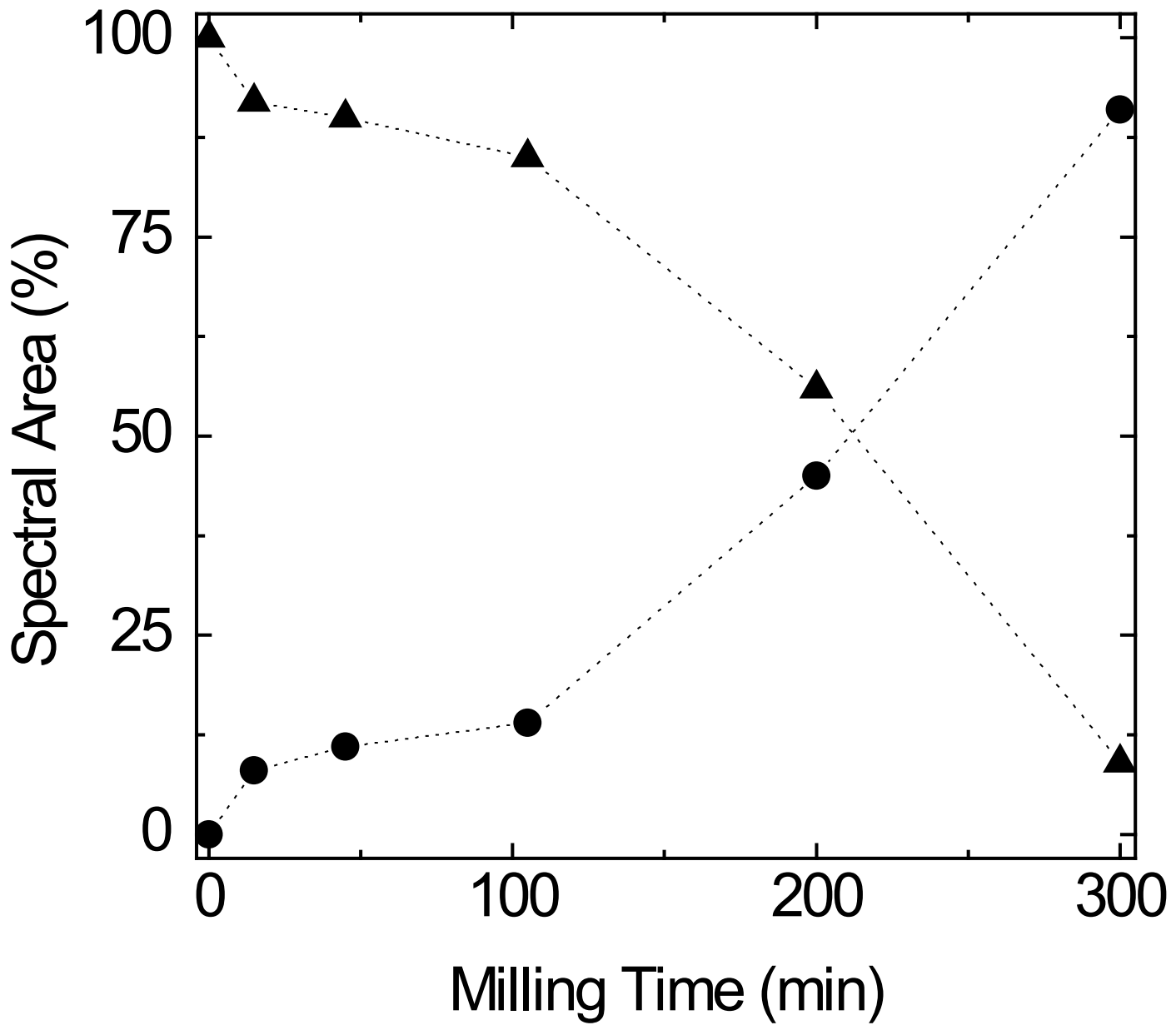

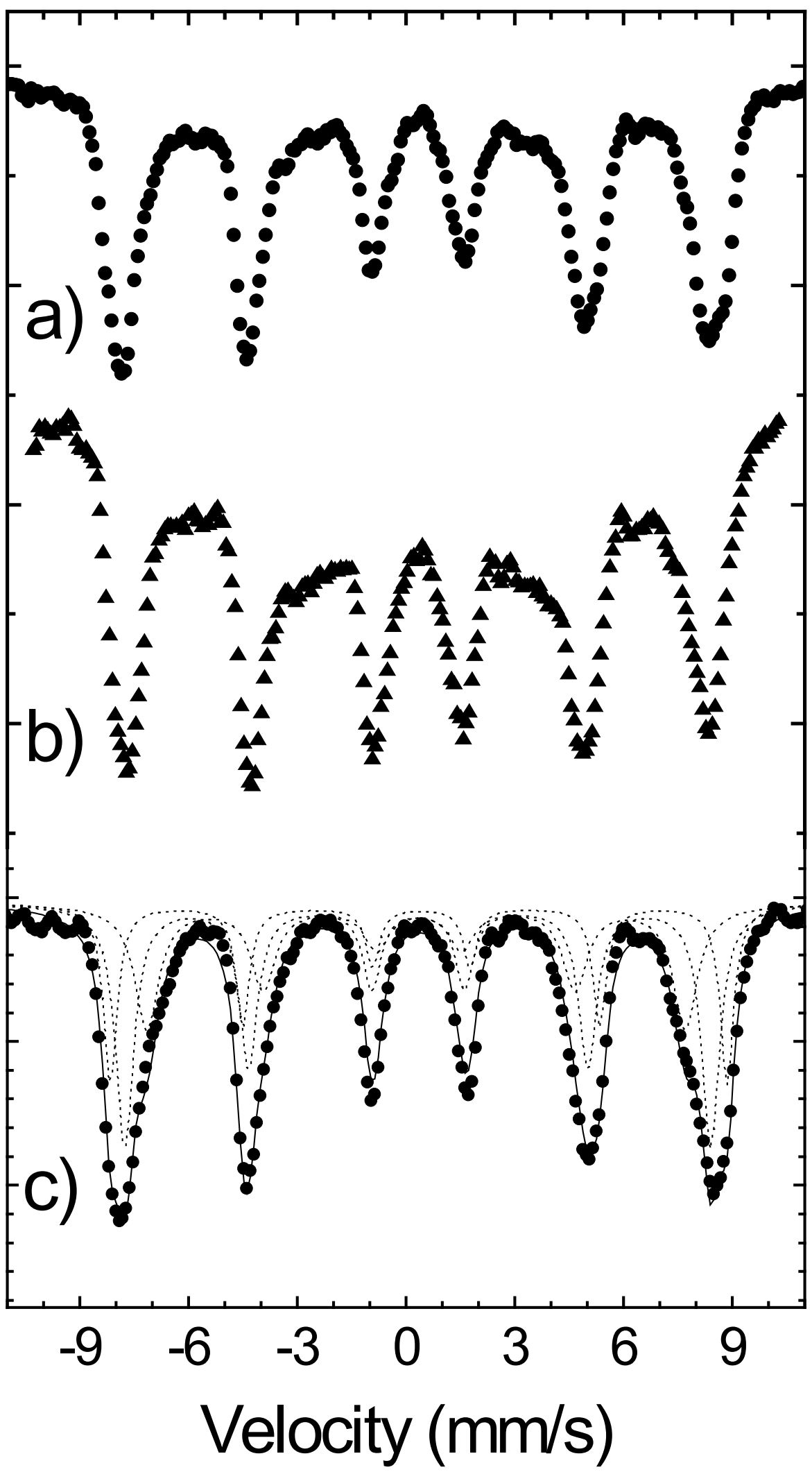

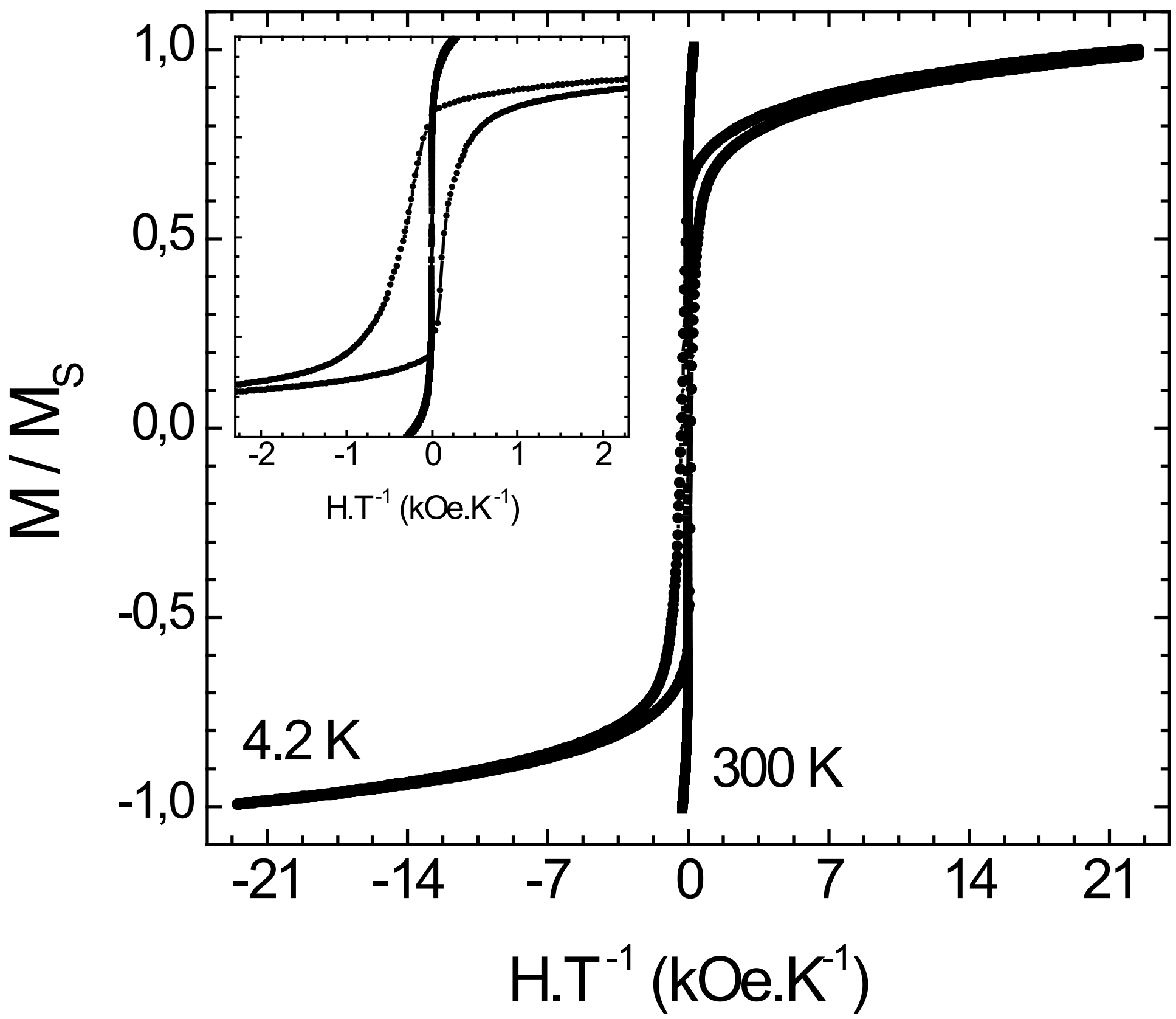